\documentclass[
superscriptaddress,
pra,
twocolumn,
showpacs,
showkeys,
aps,
floatfix,
]{revtex4-1}

\usepackage[colorlinks=true,linkcolor=blue,urlcolor=blue]{hyperref}
\usepackage[utf8]{inputenc}
\usepackage{amsmath}
\usepackage{amssymb}
\usepackage{graphicx}
\usepackage{bm,bbm}

\newcommand{\rmd}{\mathrm{d}}
\newcommand{\rme}{\mathrm{e}}
\newcommand{\rmi}{i}

\newcommand{\la}{\langle}
\newcommand{\ra}{\rangle}

\graphicspath{{eps/}}
\begin{document}

\title{Minimal dissipation model for bipartite quantum systems at finite
  temperature}

\author{Alexander Ram\'\i rez Bola\~ nos}
\affiliation{Departamento de F\'\i sica, Universidad de Guadalajara,
  Blvd. Marcelino Garc\'\i a Barragan y Calzada Ol\'\i mpica,
  C.P. 44840, Guadalajara, Jalisco, M\' exico}

\author{Héctor J. G. Moreno Carri\' on}
\affiliation{Centro de Investigaci\' on en Ciencias, Universidad Aut\' onoma 
del Estado de Morelos, Cuernavaca, Morelos, M\' exico}
\affiliation{Instituto de Ciencias F\'\i sicas, Universidad Nacional 
Aut\' onoma de M\' exico, Cuernavaca, Morelos, M\' exico}

\author{Thomas Gorin}
\affiliation{Departamento de F\'\i sica, Universidad de Guadalajara,
  Blvd. Marcelino Garc\'\i a Barragan y Calzada Ol\'\i mpica,
  C.P. 44840, Guadalajara, Jalisco, M\' exico}

\begin{abstract}
We consider the reduced dynamics in a bipartite quantum system (consisting of
a central system and an intermediate environment) coupled to a heat bath at
finite temperature. To describe this situation, in the simplest possible --
yet physically meaningful way, we introduce the ``depolarizing heat bath'' as
a new minimal dissipation model. We conjecture that at sufficiently strong 
dissipation, any other dissipation model implemented in the form of a 
Markovian quantum master equation will yield the same reduced dynamics of the
central system, as the minimal model.

To support this conjecture, we study a two-level system coupled to an
oscillator mode. For the coupling between the two parts, we consider the 
Jaynes-Cummings or a dephasing coupling, while the coupling to the heat bath 
is modeled by the quantum optical or the Caldeira-Leggett master equation 
(neglecting any direct coupling between central system and heat bath). We 
then provide ample numerical evidence, for both, model-independence and
accuracy of the depolarizing heat bath model. Alongside with our study, we
investigate different regimes, where the strong coupling condition leads to 
coherence and/or population stabilization.
\end{abstract}

\maketitle

\section{Introduction}

The concept of a generic, impartial heat bath plays an important role in 
equilibrium statistical mechanics, where it is used to keep the temperature of 
thermodynamic systems (classical or quantum) well defined and constant,
without changing the (thermodynamic) properties of these systems themselves. 
In the case of the canonical ensemble, for instance, one assumes the system to 
be in contact with a much larger second system (the environment), such that 
both systems together form an isolated composite system, treatable as a 
microcanonical ensemble. Then, in order to act as a proper heat bath, the 
coupling between system and environment and the environment itself should be 
such that (i) the dynamical properties of the system remain
unchanged and (ii) thermal relaxation still occurs~\cite{Huang87}. As a
consequence, only two quantities, temperature and coupling strength, are
sufficient to completely specify the effect of that heat bath on the system.

The description of open quantum systems and their relaxation processes is 
usually more complicated. Even in the simplest case, where a description in 
terms of a Markovian quantum master equation is possible~\cite{GoKoSu76,Lin76}, 
the required dissipation (Lindblad) terms, will depend on the dynamics of the 
system, the coupling operator(s), and the way these interact with the degrees 
of freedom of the heat bath~\cite{BrePet02}. Moreover, at the end of the 
relaxation process, the system returns to equilibrium states, which are 
thermal mixtures of the eigenstates of some modified system 
Hamiltonian~\cite{BrePet02}. This implies changes in the thermodynamic 
properties of the system, in contrast to the setting presented above. 

In order to resolve this conflicting situation, we introduce an
intermediate system, which interacts with the heat bath as postulated by an 
ideal thermal contact, while allowing any type of interaction with the central 
system. Formally, this can be done by selecting appropriately some of the 
degrees of freedom of the environment, treating them as the intermediate 
system; see for instance the Refs.~\cite{Garra97,Dal01,Rod12,Hofer17}.
We then use the ``depolarizing heat bath'' model for describing 
the effect of the external heat bath on the intermediate system, which comes as 
close as possible to the idea of an impartial thermal contact. Its high 
temperature limit has been derived in the context of random matrix models for 
decoherence~\cite{GPKS08}. In the context of quantum thermal 
machines~\cite{Skrzypczyk11,BohrNJP15,Hofer17}, this 
model is known as ``reset model''; as such it has been introduced 
in~\cite{Skrzypczyk11} for describing two-level systems coupled to thermal 
reservoirs.

We then consider dissipative bipartite systems, consisting of a two-level 
system (qubit) and a harmonic oscillator mode, which plays the role of the 
intermediate system in the setup mentioned above. We show that for different 
types of couplings (between qubit and oscillator mode) and different 
dissipation models, the reduced dynamics of the qubit always tends to the of 
the depolarizing heat bath, as the coupling to the heat bath (measured by the 
dissipation rate) is increased.

As the bipartite system we consider a central two-level system (qubit), 
coupled to a harmonic oscillator mode as ``intermediate system''. Qubit
and oscillator are coupled either by a Jaynes-Cummings (JC), or a dephasing
(D) coupling. As heat baths or dissipation models, we consider the quantum 
optical (QO) and the Caldeira-Leggett (CL) master equation. 
The JC coupling together with the QO master equation, 
represents one of the paradigmatic models in quantum optics, the dissipative 
Jaynes-Cummings model~\cite{JayCum63,BarKni86,Puri2001,Rai01}; see also the 
recent special issue, Ref.~\cite{Andrew-D-Greentree}. The 
dephasing coupling case has experimental relevance, for instance as a 
simplified description of the dynamics of defect centers taking into account 
the coupling to lattice phonons~\cite{BeToBi14}. The CL master equation finally 
is another paradigmatic model for quantum dissipation, as it provides the 
quantum analog to the classical damped harmonic oscillator in the underdamped 
regime~\cite{CalLeg83,Hu92}, applicable for instance to quantum Brownian 
motion.

The depolarizing heat bath master equation (or more precisely, its high 
temperature limit) has been derived in the context of random matrix models for 
decoherence~\cite{GPKS08}, considering the reduced dynamics of one system 
described by a random matrix ensemble, in the presence of an environment, 
described by a similar but statistically independent ensemble. It has then 
been used, to describe a dissipative bipartite system with the near 
environment modeled by random matrix theory~\cite{MGS15,GMS16RTSA}. There,
it was shown that the decoherence rate in the central system may be inversely 
proportional to the dissipation rate in the near environment. Related phenomena 
of coherence and/or population stabilization have been observed in other 
dissipative bipartite systems including the dissipative JC model: The first 
reference is probably Ref.~\cite{Cirac92}, but more explicit accounts of the 
matter can be found in Ref.~\cite{Fon08b} and very recently in 
Ref.~\cite{TorSel17}. Finally, the case of a deterministic quantum chaotic spin 
system has been studied in~\cite{GonPineda}.

The paper is organized as follows: In the next section, Sec.~\ref{M}, the
different models are introduced. They all have the same structure, a quantum
master equation for the mixed quantum state of central system and near 
environment. In Sec.~\ref{E}, we present  numerical simulations for the
different models, focussing on the accuracy of our ``depolarizing heat bath''
model, and the above mentioned stabilization effect. Finally, we present our
conclusions in Sec.~\ref{C}.

\section{\label{M} General model}

In this paper, we consider bipartite open quantum systems, strongly coupled to
an external heat bath. Typical examples are (i) the dissipative 
Jaynes-Cummings (JC) model, and (ii) the Caldeira-Leggett (CL) model coupled to
an internal degree of freedom.
We distinguish between the central system, assumed to be a two-level system
(qubit) and the much larger near environment, here chosen as a harmonic
oscillator mode with angular frequency $w_\rme$. We assume that only the near 
environment is coupled to the heat bath. The evolution of the bipartite system 
is modeled via a quantum master equation, which consists of a Hamiltonian part 
describing the dynamics of central system and near environment, and a 
dissipative part, which describes the effect of the heat bath on the near 
environment. The dissipative part, has two main control parameters, the 
temperature and the coupling strength. 

Dividing the quantum master equation in original physical units by 
$\hbar w_\rme$ and switching from physical time $t_{\rm ph}$ to the 
dimensionless time $t= w_\rme t_{\rm ph}$, we are left with a differential 
equation of the following form
\begin{align}
\dot\varrho = -\rmi\, [H_x \, ,\varrho] 
   - \frac{\kappa}{2}\; \mathcal{D}[\varrho] \; ,
\label{M:MaEc}\end{align}
where $H_x$ denotes the Hamiltonian which would govern the dynamics of 
central system and near environment, if they were perfectly isolated, and 
$\mathcal{D}[\varrho]$ denotes the dissipation term, which would model the 
effect of the heat bath on the near environment, in the absence of any 
central system. Here, we introduced the parameter $\kappa$ as a dimensionless
dissipation rate, \textit{i.e.} $\kappa = \gamma/w_\rme$, where $\gamma$ is the
physical energy dissipation rate. 

For the Hamiltonian $H_x$, the following two models will be considered:
(i) The Jaynes-Cummings model ($x = {\rm JC}$), 
\begin{align}
H_{JC} = \frac{\Delta}{2}\; \sigma_z + \hat a^\dagger\hat a + g\; 
   (\sigma_+\otimes\hat a + \sigma_-\otimes\hat a^\dagger) \; ,
\label{MH:defJC}\end{align}
where we eliminated the zero-point energy term, since it has no effect for the 
dynamics of the density matrix $\varrho$. With the energy difference in the 
two level system being denoted by $\hbar w_{\rm a}$, the dimensionless 
parameter $\Delta$ becomes $\Delta= w_{\rm a}/w_\rme$. Similarly, the coupling 
parameter $g$ is defined as $g= \Omega/(2w_\rme)$, where $\Omega$ is the 
Rabi-frequency of the original JC model. Note that in order to arrive at the 
JC model, a rotating wave approximation must be applied, which is justified if 
the level spacing in the qubit is equal or close to the boson energy of the 
harmonic oscillator. In our case, this means for the relative detuning
$\delta= \Delta - 1$, it must hold that $|\delta| \ll 1$.

(ii) The harmonic oscillator with dephasing coupling ($x= {\rm D}$) is defined
analogously, using the same units for time and energy as above. 
\begin{align}
H_{\rm D} = \frac{\Delta}{2}\; \sigma_z + \hat a^\dagger\hat a 
   + \frac{g}{\sqrt{2}}\; \sigma_z \otimes (\hat a + \hat a^\dagger) \; .
\label{MH:defD}\end{align}

\subsection{\label{MD} Quantum optical and Caldeira-Leggett master equations}

In this section, we introduce the quantum optical (QO) master equation 
alongside with the Caldeira-Leggett (CL) master equation. 
For convenience, we use the term ``quantum optical'' for the 
master equation of the dissipative Jaynes-Cummings model, even though there
exist of course many different types of quantum optical master equations.
The QO master equation is meant to describe the coupling of a single cavity 
mode to an ensemble of electromagnetic (photon) modes in thermal equilibrium. 
The coupling is assumed to occur due to imperfect mirrors in 
the cavity, often quantified by a finite quality factor or equivalently a 
finite dissipation rate. For the QO master equation, the dissipation term 
$\mathcal{D}[\varrho]$ in Eq.~(\ref{M:MaEc}) is replaced by
\begin{align}
\mathcal{D}_{\rm QO}[\varrho] &= (\bar n + 1)\, \mathcal{D}_0[\varrho] 
   + \bar n\, ( \hat a\, \hat a^\dagger\varrho 
   - 2\, \hat a^\dagger \varrho\, \hat a  + \varrho\, \hat a\, \hat a^\dagger)
\; , \label{MD:QO}\\
\mathcal{D}_0[\varrho] &= \hat a^\dagger\hat a\, \varrho 
   - 2\, \hat a\, \varrho\, \hat a^\dagger  + \varrho\, \hat a^\dagger \hat a
\; .
\label{MD:T0}\end{align}
Here, $\mathcal{D}_{\rm QO}$ is the finite temperature and $\mathcal{D}_0$ the 
zero-temperature heat bath (bare vacuum), and $\bar n$ is the average number 
of oscillator modes occupied at the given temperature. 

The CL master equation ~\cite{CalLeg83} is meant to describe quantum Brownian 
motion, \textit{i.e.} a heavy but still quantum mechanical particle in an 
harmonic potential, subject to dissipation due to frequent collisions with the 
particles of a finite temperature background gas. In this case, the model has 
a precise classical analog, which is the damped harmonic oscillator, with 
damping (or dissipation) rate $\gamma$. As in the quantum optical case,
$\kappa = \gamma/w_\rme$.

In the case of the CL model, the dissipation term is usually written in terms
of physical position and momentum operators, while the temperature enters the
expression via a diffusion constant. However, for the sake of a
consistent description, we rewrite the CL dissipation term, using the same 
adimensional quantities as in the QO case. The details can be found in 
App.~\ref{aC} with the result given in Eq.~(\ref{aC:CalLegDiss}) as reproduced 
here.
\begin{align}
\mathcal{D}_{\rm CL}[\varrho] = 2\rmi\, [\hat x, \{ \hat p,\varrho\} ]
   +2\, (2\bar n +1)\, [\hat x, [\hat x,\varrho]] \; ,
\label{MD:CL}\end{align}
where the dimensionless position and momentum operators $\hat x$ and $\hat p$
are defined such that $\hat a = (\hat x + \rmi\hat p)/\sqrt{2}$. 
As usual, we denote the commutator (anti-commutator) between two
operators $A,B$ as $[A,B] = AB - BA$ ($\{A,B\} = AB + BA$). 

The CL dissipation model is derived in the limit of high temperature, and the 
resulting master equation is not of Lindblad form. It may thus violate the 
positivity of the evolving density matrix~\cite{Sandulescu87,Diosi93}. However, 
as long as the temperature is not very small this problem has no significant 
effect on physical quantities~\cite{Ram09}.

In both models, $\kappa$ is the dissipation rate in units of the angular 
frequency $w_\rme$ of the oscillator mode. It describes the rate of energy 
loss, when the initial state has higher thermal energy than the respective 
heat bath; see Fig.~\ref{f:EJ:Equilib}.

\subsection{\label{MH} Depolarizing heat bath}

We are interested in relaxation processes, where only the central system is
taken out of equilibrium. Thus, we assume that the initial state of our 
bipartite system is a product state of an arbitrary initial state 
$\varrho_{\rm a}(0)$ of the central system and a thermal equilibrium state 
$w_T$ of the near environments, who's temperature $T$ agrees with that of the 
heat bath (see App.~\ref{aT}). Without coupling to the central system, that 
state would then be a equilibrium solution of the master equation. We then 
concentrate on the reduced dynamics of the central system in the regime of 
strong coupling between intermediate system and heat bath. In such a 
situation, we may assume that the details of the dynamics in the intermediate 
system are not so important than the fact that the intermediate system has a 
strong tendency to quickly return to the equilibrium state. 

To describe this situation, we consider the simplest possible quantum 
operation, which maps any mixed state directly to the equilibrium state (we 
call this operation ``depolarizing channel''). This operation is turned into a 
dissipation term in the master equation by using Milburn's theory~\cite{Mil87}.
Physically, this means that the depolarizing channel is applied to the system 
with a certain rate, $\gamma_{\rm P}$, which plays the role of the coupling 
parameter between near environment and heat bath. Comparing the resulting 
master equation with that in Eq.~(\ref{M:MaEc}), we find again that 
$\kappa = \gamma_{\rm P}/w_\rme$ and finally the following dissipation term:
\begin{align}
\mathcal{D}_{\rm DH}[\varrho] &= 2\, (\varrho 
   - {\rm tr}_\rme \varrho \otimes w_{\rm T})\; ,
\label{MD:DH}\end{align}
where $w_{\rm T}$ is the finite temperature equilibrium state of the near 
environment alone. Unfortunately, $\gamma_{\rm P}$ cannot be directly compared
to the energy dissipation rate $\gamma$, as defined in the previous models. It
is therefore not clear, how to choose $\gamma_{\rm P}$ as compared to $\gamma$
such that the previous models really converge to this DH model, at strong
coupling. According to Fig.~\ref{f:EJ:Equilib}, where we study directly the 
energy dissipation, it seems that $\gamma_{\rm P}$ and $\gamma$ should simply 
be equal. However, in subsequent studies concentrating on the reduced dynamics 
of the central system, it appears that $\gamma_{\rm P}$ should rather be equal 
to $\gamma/2$.

In order to appreciate the impartiality of the DH model, consider the dynamics 
of the oscillator mode without coupling to the qubit. 
In that case, the solution to the master equation~(\ref{M:MaEc}) 
with dissipation term $\mathcal{D}_{\rm DH}$ is of the form 
$\varrho_{\rm a}(t)\otimes \varrho_{\rm e}(t)$, with
\begin{equation}
\frac{\rmd}{\rmd t}\, \tilde\varrho_\rme(t) = -\kappa\, \big (\,
   \tilde\varrho_\rme(t) - w_T\, \big ) \; ,
\end{equation}
where $\tilde\varrho_\rme(t)$ describes the cavity state in the interaction
picture. According to this equation, all matrix elements of 
$\tilde\varrho_\rme(t)$ converge exponentially towards the matrix elements of 
the thermal equilibrium state $w_T$, with the same rate $\kappa$.

\subsection{\label{ME} Evolution}

We are mainly interested in the evolution of the qubit (the central system),
under the coupling to the cavity mode and the external heat bath as a 
composite environment. We compute the evolution by numerically solving the
master equation~(\ref{M:MaEc}) with the help of a standard solver for ordinary
differential equations. Typically, we use about 20 up to 40 basis states in
the Hilbert space of the oscillator. Choosing an initial state of the form
$\varrho(0) = \varrho_{\rm a}(0)\otimes w_T$, we obtain the evolution of the 
full bipartite system as $\varrho(t)$. From this quantity, we compute the 
state of the two-level system by taking a partial trace over the oscillator 
mode.

\section{\label{E} Numerical Results}

In what follows, we consider the evolution of the qubit under different
couplings (Jaynes-Cummings, dephasing) and different dissipation models
(quantum-optical, Caldeira-Leggett, and depolarizing heat bath) as introduced 
in Eqs.~(\ref{MD:QO}) and~(\ref{MD:DH}). We concentrate on the strong coupling 
regime, where we expect the depolarizing heat bath (DH) model to be a good 
substitute for any other model. Thus, the main objective in this section 
consists in finding numerical evidence that the state evolution in all cases 
converge to the evolution under the DH model, as $\kappa/g \to \infty$. Note 
that we expect this to be valid only if the initial state of the near 
environment is equal (or at least sufficiently close) to the thermal 
equilibrium state to be reached at the end of the relaxation process.
We solve the master equation numerically as a system of linear 
differential equations using the eigenstates of the harmonic oscillator mode
as an orthonormal basis to expand all operators involved. Unless states 
otherwise, the basis is limitted to quantum numbers $n \le n_{\rm max} = 40$.

The second objective addresses the stabilization effect found in some related
models: (i) In a random matrix model for dephasing 
coupling~\cite{MGS15,GMS16RTSA}, we found the stabilization effect only for
$g \ll 1$ and $\kappa/g \gg 1$. On this basis, we could for instance exclude
the quantum Zeno effect as a possible explanation for the effect. (ii) For the
dissipative JC model~\cite{Cirac92,Fon08b,TorSel17} (JC coupling between qubit
and oscillator mode, and quantum optical master equation), the only requirement
is $\kappa/g \gg 1$. 
To these cases, we will add a third: (iii) The harmonic oscillator mode with 
dephasing coupling to a qubit and quantum optical dissipation, a model which
is quite similar to the one considered in Ref.~\cite{BeToBi14}. As we will see 
below, in this case the only requirement is $\kappa \gg 1$.

\subsection{\label{ER} Dissipation rate}

\begin{figure}
\includegraphics[width=0.5\textwidth]{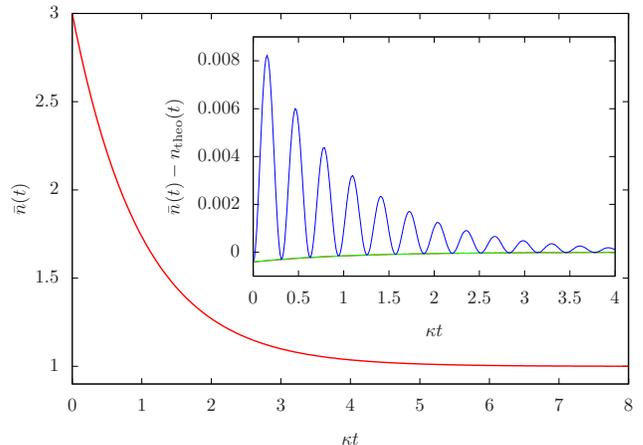}
\caption{(Color online) Equilibration without qubit ($g=0$) for 
$\kappa = 0.1$ and a initial thermal state with $\bar n(0) = 3$, for the 
quantum optical (QO) dissipation model. The inset shows the difference between 
the different dissipation models and the theorical model in 
Eq.~(\ref{ED:Equilib}): Red line ($\mathcal{D}_{\rm QO}[\varrho]$, quantum 
optical); blue line ($\mathcal{D}_{\rm CL}[\varrho]$, Caldeira-Leggett);
green line ($\mathcal{D}_{\rm DH}[\varrho]$, depolarizing heat bath).}
\label{f:EJ:Equilib}\end{figure}

We would like to make sure that the dissipation rate, controlled in the 
master equation~(\ref{M:MaEc}) by the parameter $\kappa$, is the same for 
all dissipation models, considered. For that reason, we analyze the 
equilibration of the oscillator mode without coupling to the 
qubit. We choose the initial state to be a thermal equilibrium state with 
$\bar n(0) = 3$ and compute its evolution and average energy 
$\bar n(t)= \la\hat a^\dagger \hat a\ra_{\varrho(t)}$ when the 
temperature of the heat bath is such that $\bar n_{\rm eq} = 1$. 

The simplest theoretical expectation for $\bar n(t)$ is an 
exponential decay towards the new equilibrium energy, i.e.
\begin{align}
   n_{\rm theo}(t) = \bar n_{\rm eq} + [\, \bar n(0) - \bar n_{\rm eq}\, ]\; 
   \rme^{-\kappa t}\; .
\label{ED:Equilib}\end{align}
This expectation is verified in Fig.~\ref{f:EJ:Equilib}. In the 
main panel, we show the behavior of $\bar n(t)$ for the quantum optical 
dissipation model (red solid line); in the inset we show the difference between 
all three dissipation models and the theoretical expectation, 
Eq.~(\ref{ED:Equilib}). In that graph, the quantum optical (QO) and the 
depolarizing heat bath (DH) cases are lying almost on top of each other, 
so that only the green curve can be seen. The Caldeira-Leggett model (blue 
line) shows clear deviations of the order of one percent. The QO and DH cases 
show a small deviation, noticeable at $\kappa t$ close to zero, only. We 
attribute this to the finite number of basis states 
($0 \le n \le n_{\rm max} = 40$) we have been using for the simulations. The 
differences in the CL result in the form of damped oscillations, probably are
an artefact, related to the Caldeira-Leggett master equation being valid at 
large temperatures, only.

\subsection{\label{EJ} Jaynes-Cummings coupling at zero temperature}

For zero and non-zero temperature, the only relevant energy (frequency) scale 
is given by $g$. This is due to the fact that the Jaynes-Cummings Hamiltonian
in Eq.~(\ref{MH:defJC}) can be decomposed into two commuting parts, one of 
which is an observable for the number of excitations in the system: 
$H_{\rm ext} =  \sigma_z/2 + \hat a^\dagger\hat a$. Therefore, if 
time-dependent quantities are plotted against $g\, t$, the only independent 
parameters left are the detuning, $(\Delta - 1)/g$, and the relative coupling
strength to the external heat bath, $\kappa/g$. This is true for all 
dissipation models discussed in this paper, as given in the 
Eqs.~(\ref{MD:QO}) --~(\ref{MD:DH}).

\begin{figure}
\includegraphics[width=0.5\textwidth]{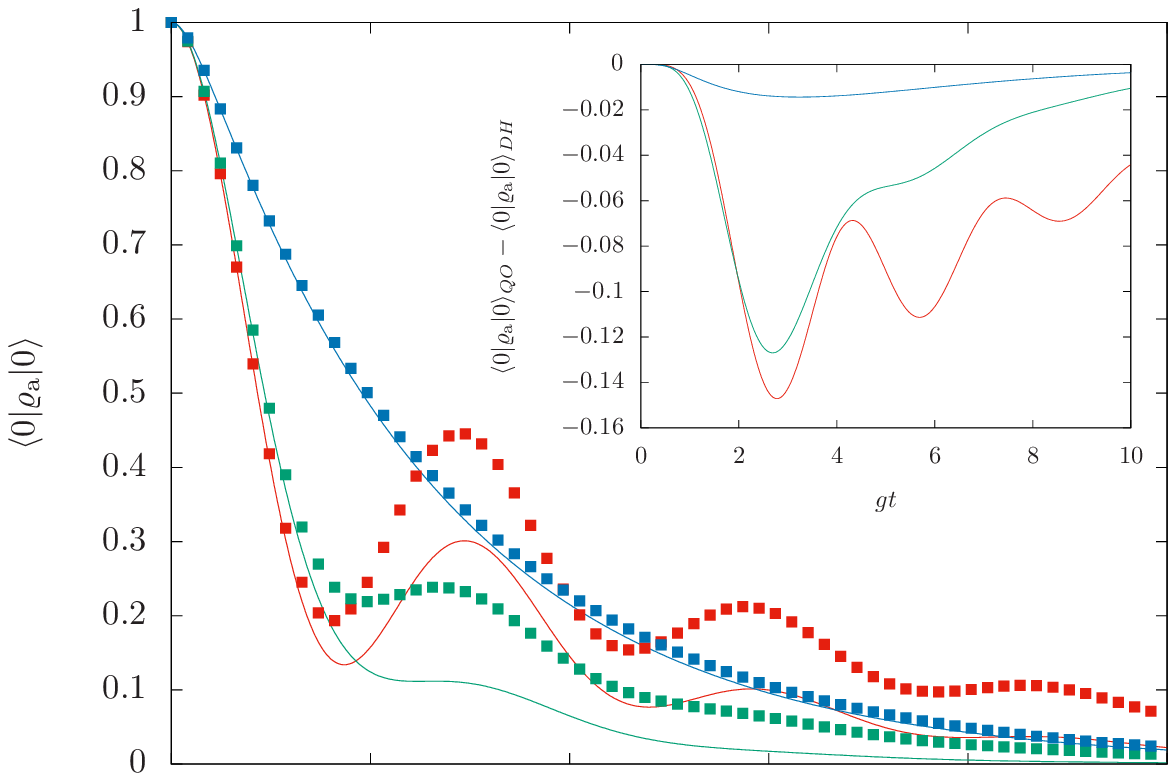}

\vspace{-5mm}

\includegraphics[width=0.5\textwidth]{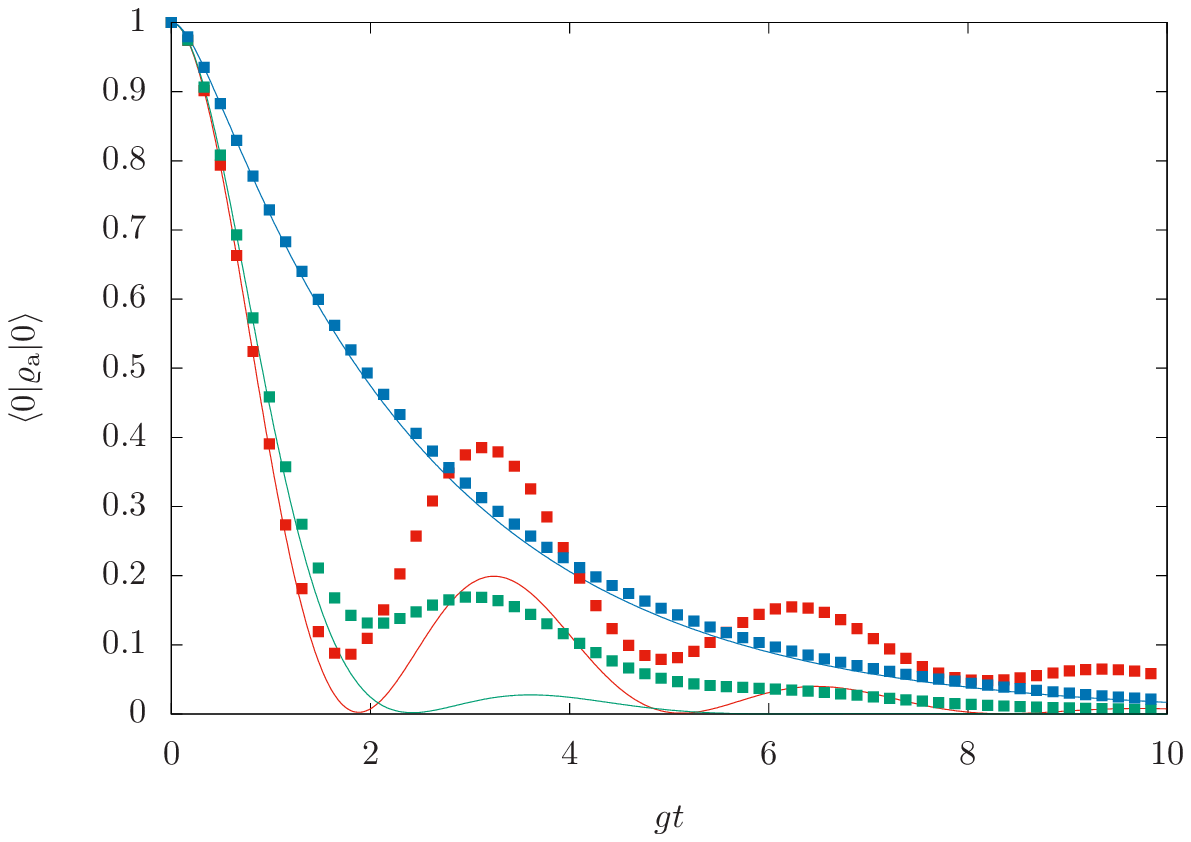}
\caption{(Color online) Exited state population as a function of scaled time
$gt$ for the dissipative JC model at zero temperature. Comparison between the
original quantum optical dissipation (solid lines) and the depolarizing heat
bath model (square points). The color coding identifies the different 
dissipation rates: $\kappa/g = 1.0$ (red), $2.0$ (green), $10.0$ (blue). 
For the DH model, all rates have been reduced by one half. The upper 
panel shows the large detuning case, $(\Delta - 1)/g= 0.8$; in the lower panel 
$(\Delta - 1)/g= 0.1$.}
\label{f:EJ:TorSel}\end{figure}

Let us first consider the behavior of the excited state population; this is 
done in Fig.~\ref{f:EJ:TorSel}. There, we choose 
$\varrho_{\rm a}(0) = |0\ra\la 0|$ and select two different 
values for the detuning, $(\Delta - 1)/g = 0.8$ (upper panel) and $0.1$ (lower 
panel), where the first case is identical to a case treated in 
Ref.~\cite{TorSel17} (Fig.~2 in that reference). We compare the behavior of 
the excited state population $\la 0|\varrho_{\rm a}|0\ra$, under the quantum 
optical (solid lines) and the depolarizing heat bath (square points) 
dissipation model. For $\kappa$ of the order of $g$, 
$\la 0|\varrho_{\rm a}|0\ra$ shows rather strong oscillations, which are 
reproduced qualitatively by the DH model, but quantitative differences remain. 
Since the temperature was chosen to be zero, the population always tends to 
zero at large times.

As expected, we find that both models lead to the same behavior if the 
coupling to the heat bath is sufficiently strong. Unexpectedly however, in 
order to achieve that, we had to reduce the dissipation rate for the DH model 
by a factor of two as compared to the QO model. For the large detuning case,
we show the difference between the two dissipation models in the inset of the
upper panel, providing clear evidence of the behavior just described. 
Finally, we find the expected stabilization effect, without much differences 
between large and small detuning. 

\begin{figure}
\includegraphics[width=0.5\textwidth]{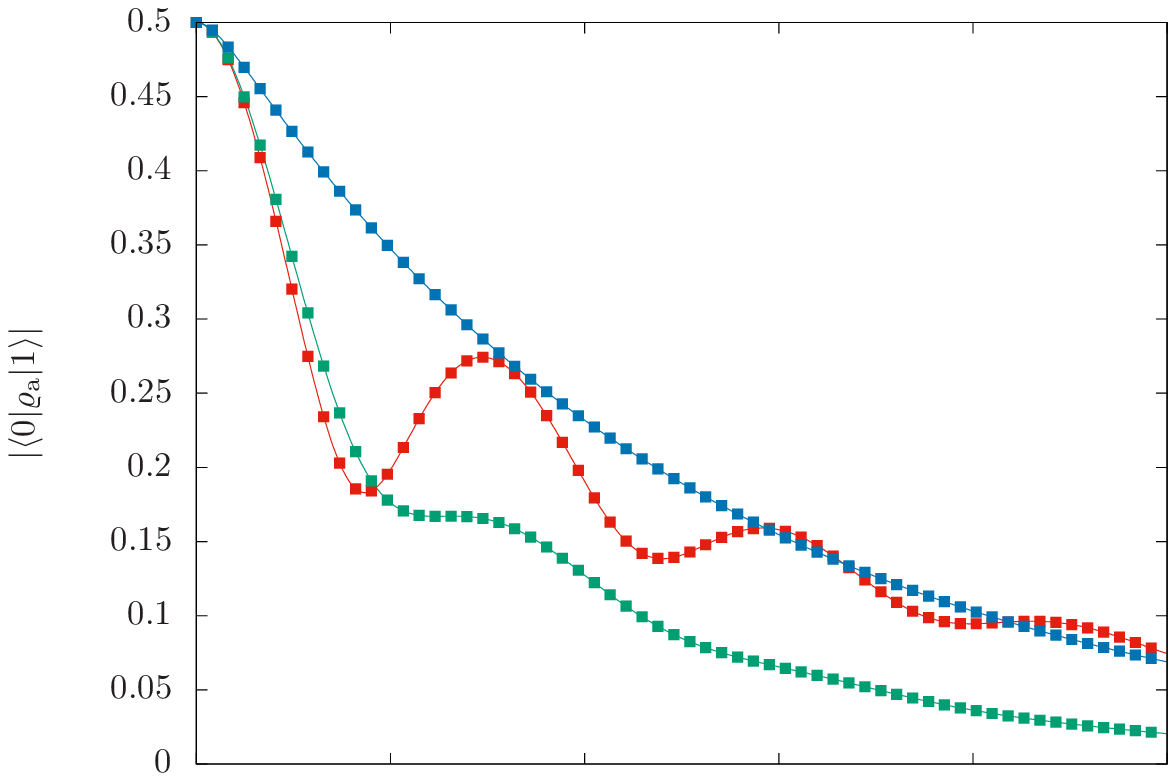}

\vspace{-5mm}

\includegraphics[width=0.5\textwidth]{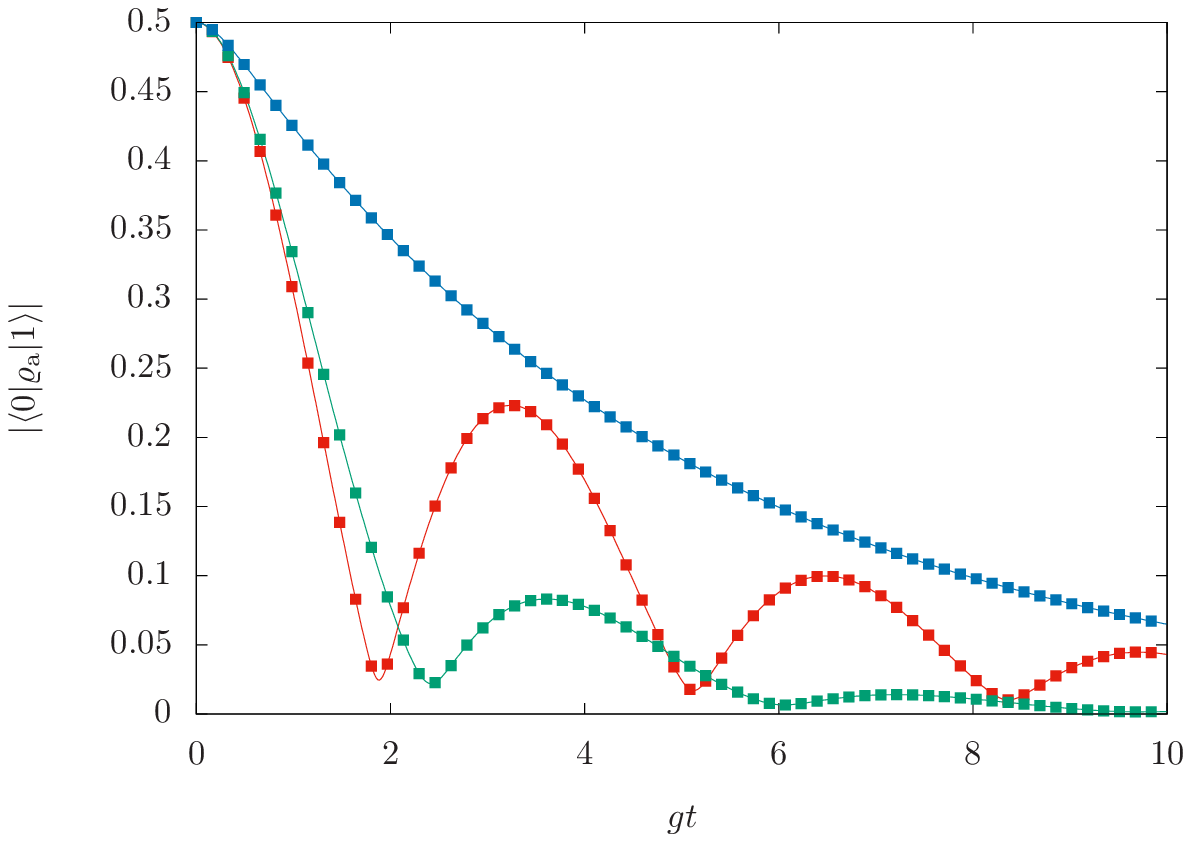}
\caption{(Color online) The coherence for an eigenstate of $\sigma_x$ as 
initial state under exactly the same conditions as in Fig.~\ref{f:EJ:TorSel}. 
In this case, the QO model (solid lines) and the DH model (square points)
yield exactly the same result.}
\label{f:EJ:TorSel-nondiagT0}\end{figure}

Fig.~\ref{f:EJ:TorSel-nondiagT0} shows the behavior of the coherence (absolute
value of the non-diagonal element of the qubit state), when the initial state 
is an eigenstate of $\sigma_x$,
\begin{align}
\varrho_{\rm a}(0) = \frac{1}{2} 
   \begin{pmatrix} 1 & 1 \\ 1 & 1 \end{pmatrix}\; .
\label{EJ:sigxestate}\end{align}
In this case, the dynamics is exactly the same for both dissipation models
(this is no longer true, when the temperature is different from zero). Again,
we find a stabilization (slower decay of the coherence), when the coupling to
the heat bath is increased.

To summarize, for zero temperature and Jaynes-Cummings coupling, the 
QO-dynamics converges to the DH-dynamics for the populations (diagonal 
elements of the reduced qubit state), for the coherence (nondiagonal elements)
both models yield exactly the same behavior. We also find the expected 
stabilization effect at $\kappa > g$ for both, for the diagonal as well as for
the non-diagonal elements. In all cases, $\kappa$ had to be divided by two in 
the DH model, in order to reach agreement at strong coupling.

\subsection{\label{EF} Jaynes-Cummings coupling at finite temperature}

In this section, we compare all three dissipation models, the 
quantum optical (QO), the depolarizing heat bath (DH) and the Caldeira-Leggett 
model (CL). For the CL model, the angular frequency $w_\rme$ of the oscillator 
mode provides an additional energy scale. Since the model is valid in the 
underdamped case only (in terms of the classical damped harmonic oscillator),
we should not consider the CL model at dissipation rates beyond $\kappa =1$.

\paragraph*{Thermal equilibrium in the central system}
Besides the comparison of the different dissipation models, and the question 
about the stabilization effect, we may ask whether the coupling between qubit 
and cavity mode may be considered as a thermal contact. In this case, we would 
expect that the final state of the bipartite system is a product state with 
the cavity mode and the qubit in thermal equilibrium states corresponding to 
the same temperature $T$, established by the external heat bath.

Since we quantify the temperature in terms of the average number of excited 
modes $\bar n = \la\hat a^\dagger\hat a\ra$ at thermal equilibrium (see 
App.~\ref{aT}). Let us assume that the two-level atom (qubit) reaches an 
equilibrium state, close to the thermal equilibrium state. According to 
Eq.~(\ref{aT:HOtempbarn}), its adimensional inverse temperature is given by
\begin{align}
b &= \frac{\hbar w_\rme}{k_{\rm B} T} = 2\, {\rm atanh}[ (2\bar n +1)^{-1}] 
\notag\\
&\Rightarrow\quad \la 0|\varrho_a^{\rm eq}|0\ra 
   = \frac{1 - \tanh(\Delta b/2)}{2}\; .
\label{EF:reqnbar}\end{align}

\begin{figure}
\includegraphics[width=0.5\textwidth]{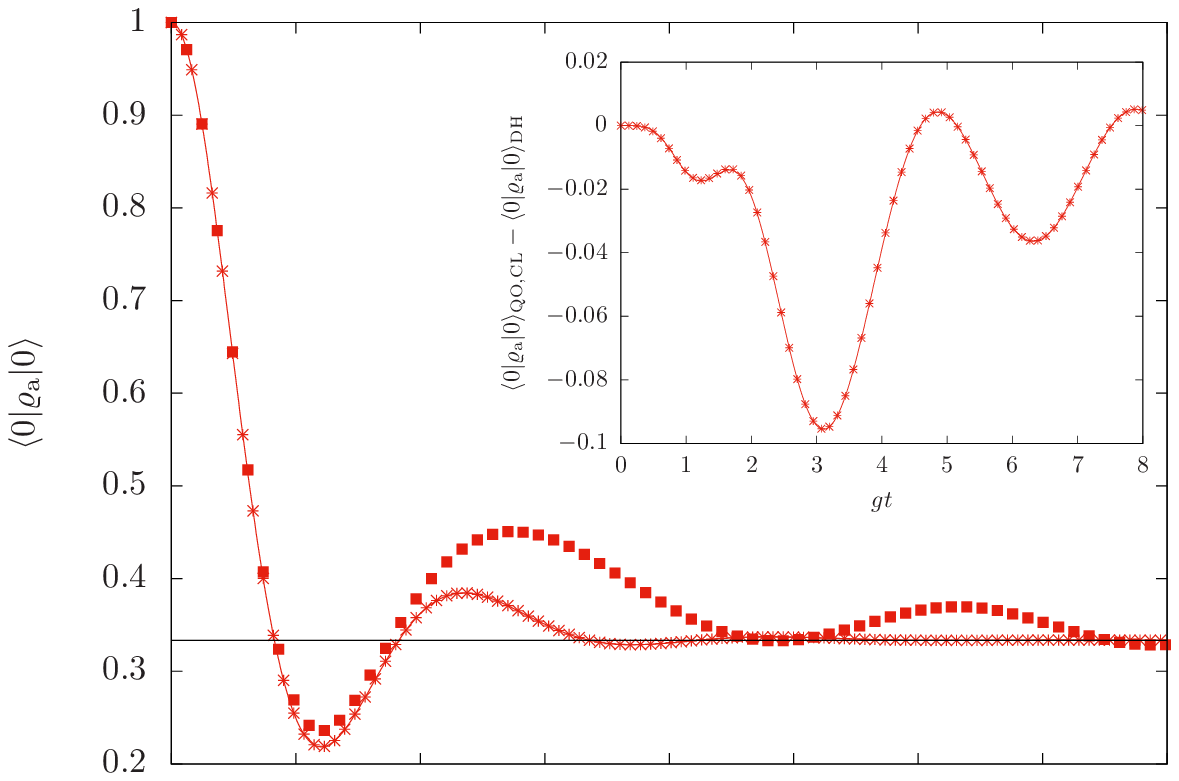}

\vspace{-5mm}

\includegraphics[width=0.5\textwidth]{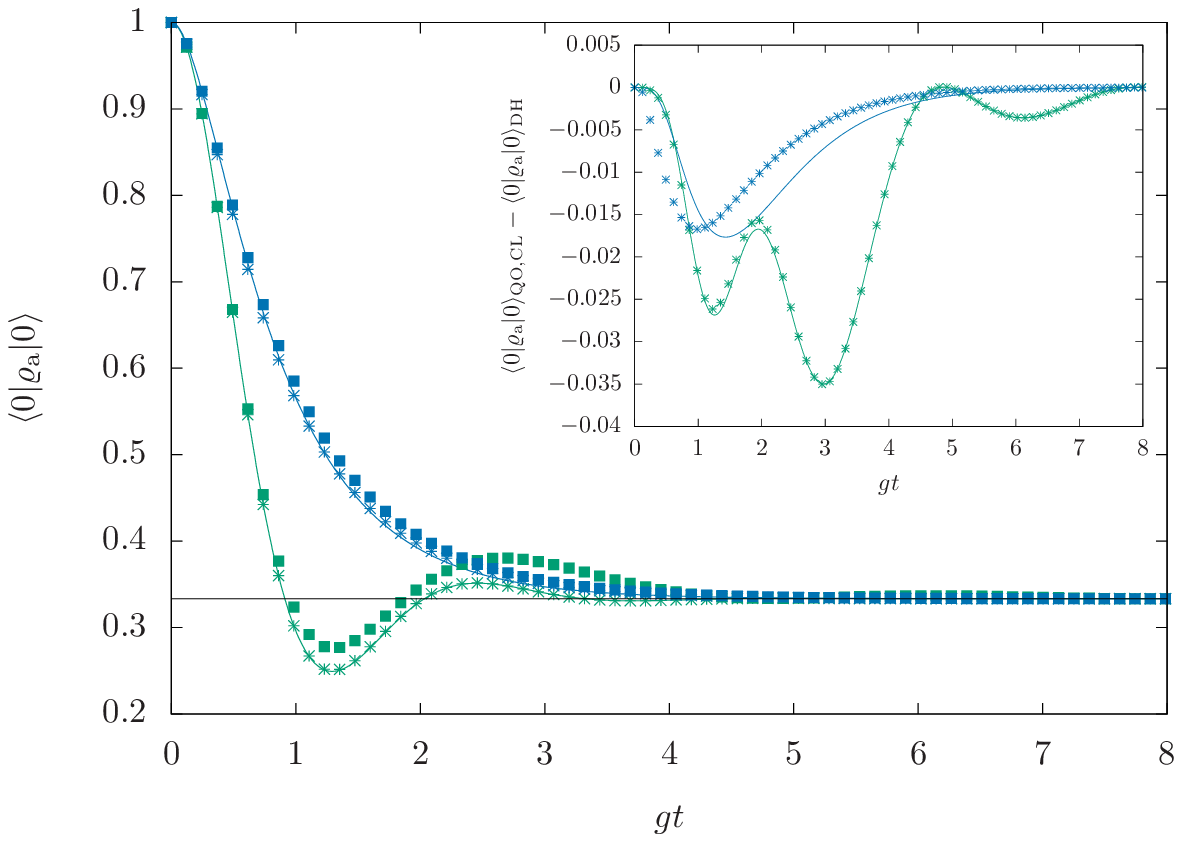}
\caption{(Color online) Exited state population as a function of scaled time
$gt$ for the dissipative JC model for small detuning $(\Delta - 1)/g= 0.1$ and
finite temperature, $\bar n = 1.0$. The horizontal black line shows the thermal 
equilibrium value expected on the basis of Eq.~(\ref{EF:reqnbar}). Solid lines
show results for the QO model, asterisks those for the Caldeira-Leggett model,
and square points those for the DH model. The color 
coding identifies the different dissipation rates: $\kappa/g = 1.0$ (red; upper
panel), $2.0$ (green; lower panel) and $10.0$ (blue; lower panel). 
Again, for the DH model, all rates have been reduced by one half. In the two
insets, we plot the difference between the results for both models. }
\label{f:EF:TempDep}\end{figure}

In Fig.~\ref{f:EF:TempDep}, we analyze the same cases as in 
Fig.~\ref{f:EJ:TorSel} but at finite temperature, $\bar n = 1$. This allows us
to include results for the Caldeira-Leggett (CL) dissipation model, also. Note
however, that we performed all calculations with $g= 0.1$, such that for the 
CL model the permissible values for $\kappa/g$ are limited to 
$\kappa/g \le 10$. To our surprise, the CL results are practically 
indistinguishable from the QO results in this case, except for the strongest
dissipation, $\kappa/g = 10$, where we find a small difference (see inset of
the lower panel).

For clarity, Fig.~\ref{f:EF:TempDep} shows results for the small detuning case 
$(\Delta - 1)/g = 0.1$, only. This allows us to plot the curves for 
$\kappa/g = 1.0$ in the upper panel and all others in the lower panel. The 
results for large detuning, $(\Delta - 1)/g = 0.8$ (not shown), are very 
similar.

Qualitatively, we find a very similar behavior as in the zero temperature
case: A decay to the equilibrium value, and superimposed oscillations, as long 
as the dissipation rate $\kappa$ is not too large. Again we observe a 
stabilization effect, when the dissipation rate is increased. And again, the QO and CL results tend to converge to the DH results (at half the dissipation 
rate) in the limit of strong dissipation. 

In Fig.~\ref{f:EF:TempDep}, the equilibrium state is no longer the ground 
state $|1\ra$, but a thermal mixture between excited and ground state. The 
solid black horizontal line, shows the probability of the qubit to be in the 
excited state, provided it is in a thermal state, in thermal equilibrium with 
the cavity mode, according to Eq.~(\ref{EF:reqnbar}). The figure clearly 
provides evidence that the JC coupling between qubit and oscillator mode, acts
indeed as a thermal contact, which forces the qubit into a thermal equilibrium
state at the same temperature than external heat bath and oscillator mode.

To conclude this section about the Jaynes-Cummings model at finite temperature,
we consider again the behavior of the coherence. In the two remaining figures 
to be shown, we remove the CL dissipation case, since in the region where it
is valid ($\kappa \le 1$), the results are practically indistinguishable from 
the QO case.

\begin{figure}
\includegraphics[width=0.5\textwidth]{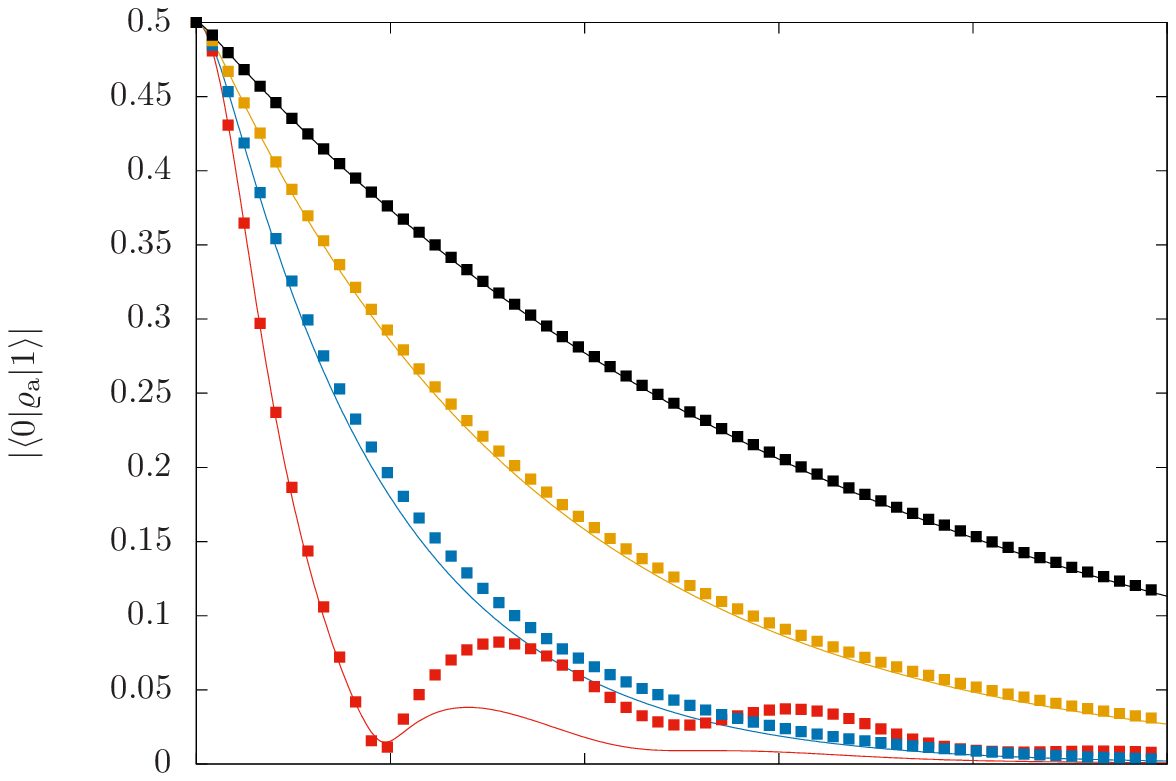}

\vspace{-5mm}

\includegraphics[width=0.5\textwidth]{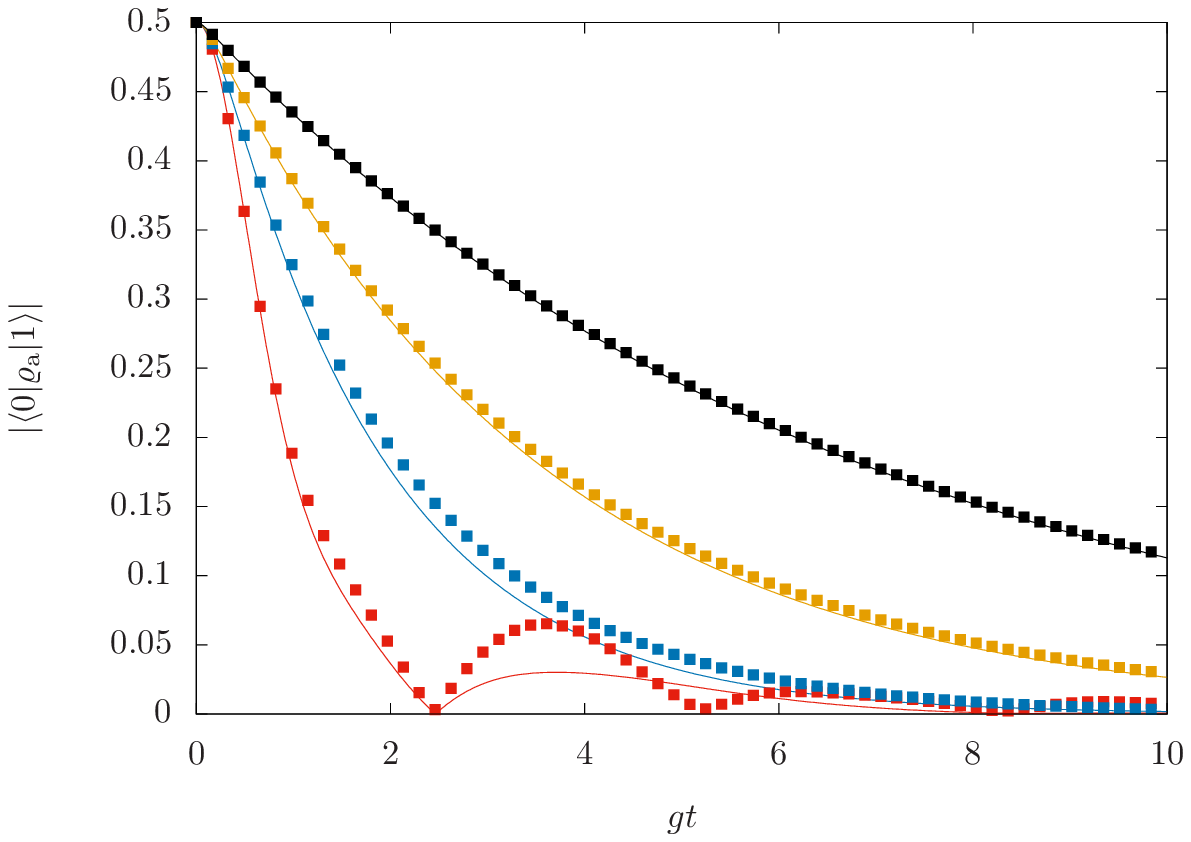}
\caption{(Color online) Coherence for an eigenstate of $\sigma_x$ as 
initial state at finite temperature $\bar n = 1.0$. Solid lines show results 
for the QO model, square points those for the DH model. The color coding 
identifies the different dissipation rates: $\kappa/g = 1.0$ (red), $10.0$
(blue), $20$ (yellow), and $40$ (black). For the DH model, all rates have been 
reduced by one half. The upper (lower) panel show the case of large, 
$(\Delta - 1)/g= 0.8$ (small, $(\Delta - 1)/g= 0.1$) detuning. }
\label{f:EF:TorSel-nondiag}\end{figure}

\begin{figure}
\includegraphics[width=0.5\textwidth]{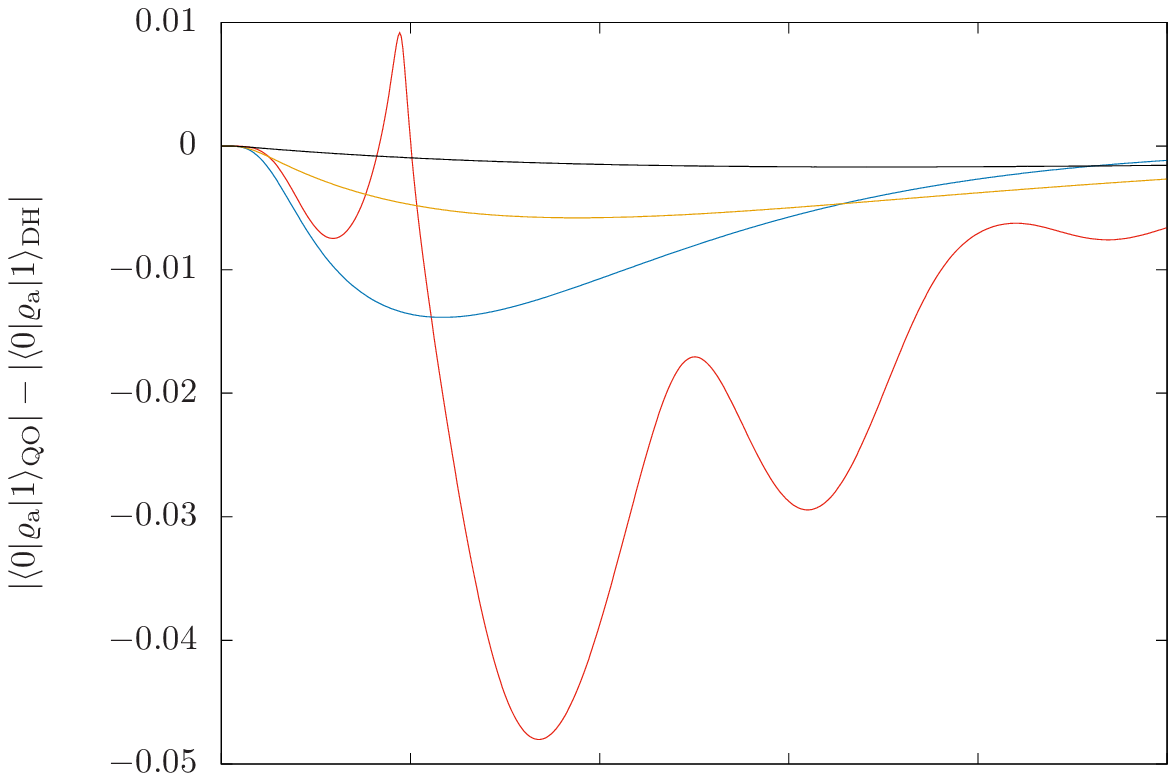}

\vspace{-5mm}

\includegraphics[width=0.5\textwidth]{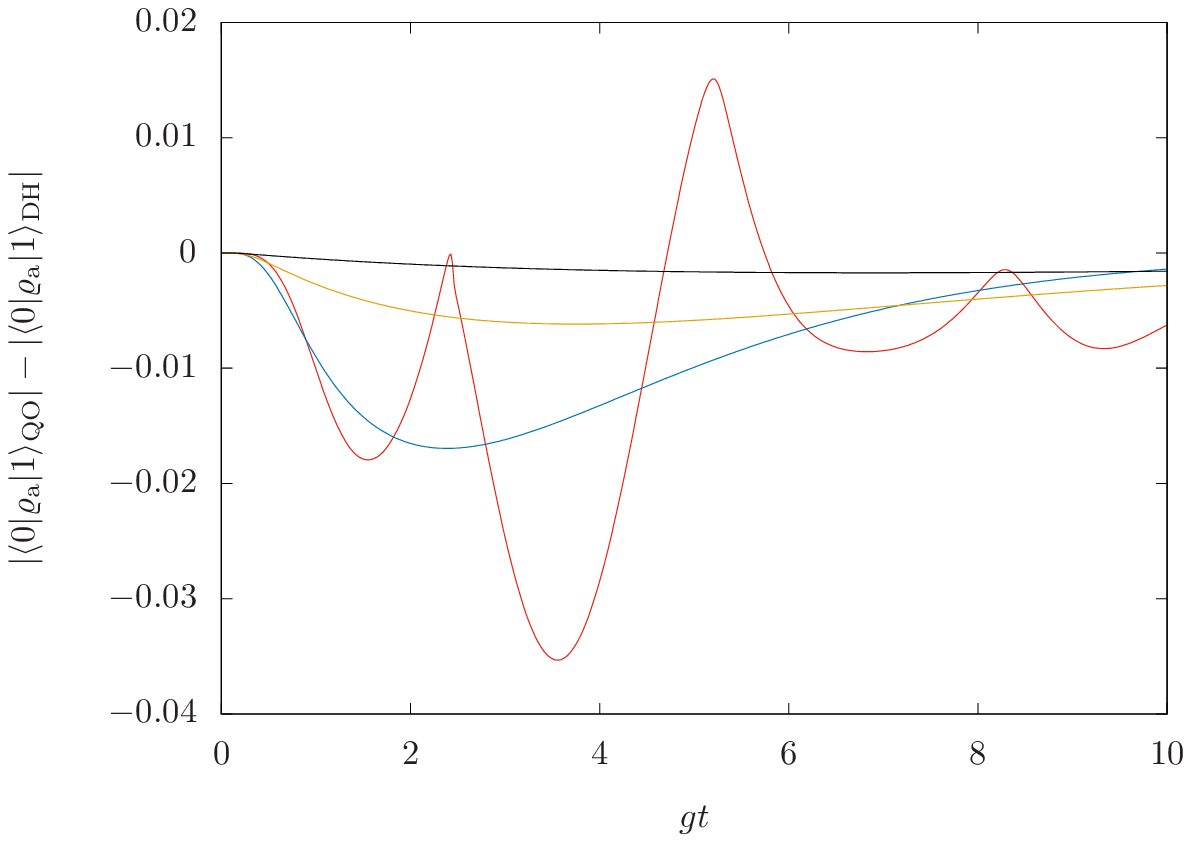}
\caption{(Color online) Differences between the 
curves of the Fig.~(\ref{f:EF:TorSel-nondiag}), using the color coding from
there. }
\label{f:EJ:TorSel-nondiag-diff}\end{figure}

Fig.~\ref{f:EF:TorSel-nondiag} shows the coherence, for an eigenstate of 
$\sigma_x$ as initial state, as defined in Eq.~(\ref{EJ:sigxestate}). In 
distinction to the zero-temperature case, the DH result is no
longer equal to the QO result, but they become equal as $\kappa/g$ becomes
large. The stabilization effect works as efficient as in the zero temperature
case (Fig.~\ref{f:EJ:TorSel-nondiagT0}).
In Fig.~\ref{f:EJ:TorSel-nondiag-diff}, we show the differences between the 
QO and DH results from Fig.~(\ref{f:EF:TorSel-nondiag}). Again, we find the 
expected convergence to the DH model for strong coupling.

\subsection{\label{ED} Dephasing model at finite temperature}

In this section, we replace the JC coupling between the qubit and the cavity
mode by the dephasing coupling, Eq.~(\ref{MH:defD}), where the diagonal matrix 
elements of the qubit state remain constant. Without external heat bath, the 
coherence (non-diagonal element of the atom state) measures the fidelity 
amplitude for the Hamiltonian of the cavity mode, perturbed by the cavity-term
of the dephasing coupling~\cite{GCZ97,GPSS04}.

For random matrix models, the stabilization effect for the coherence has been 
demonstrated in Refs.~\cite{MGS15,GMS16RTSA}. There, it was found that the 
stabilization effect is efficient for small dephasing couplings, only, 
\textit{i.e.} $2\pi\, g$ must be small as compared to the level spacing in the 
near environment (perturbative regime). In that case, the dissipation due to
the heat bath suppresses the decoherence, just as in the cases shown here. 
For larger couplings $g$, the fidelity decays faster, eventually faster than
the Heisenberg time of the near environment. Then, there is not enough 
time for the dynamics to ``realize'' that the system is finite (discrete 
spectrum), which avoids any effect when adding dissipation. This becomes 
understandable, if one interprets the increase of the coupling as a way to
enlarge the near environment.

\begin{figure}
\includegraphics[width=0.5\textwidth]{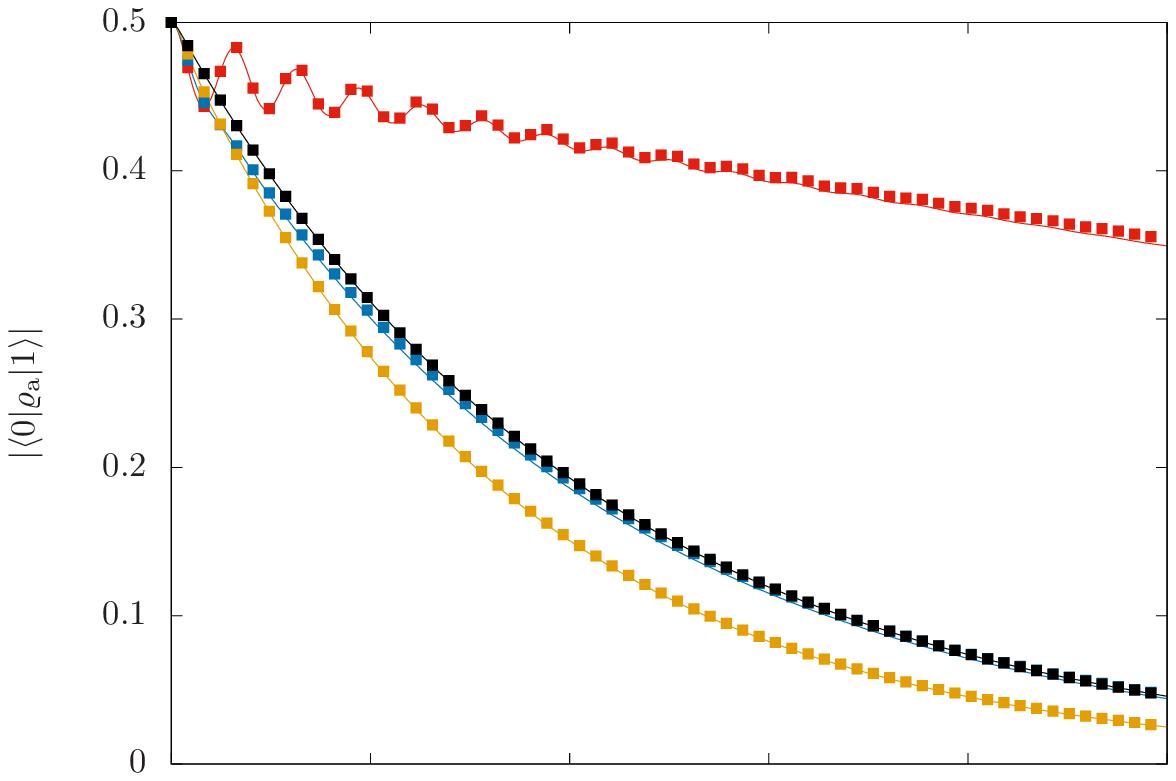}

\vspace{-5mm}

\includegraphics[width=0.5\textwidth]{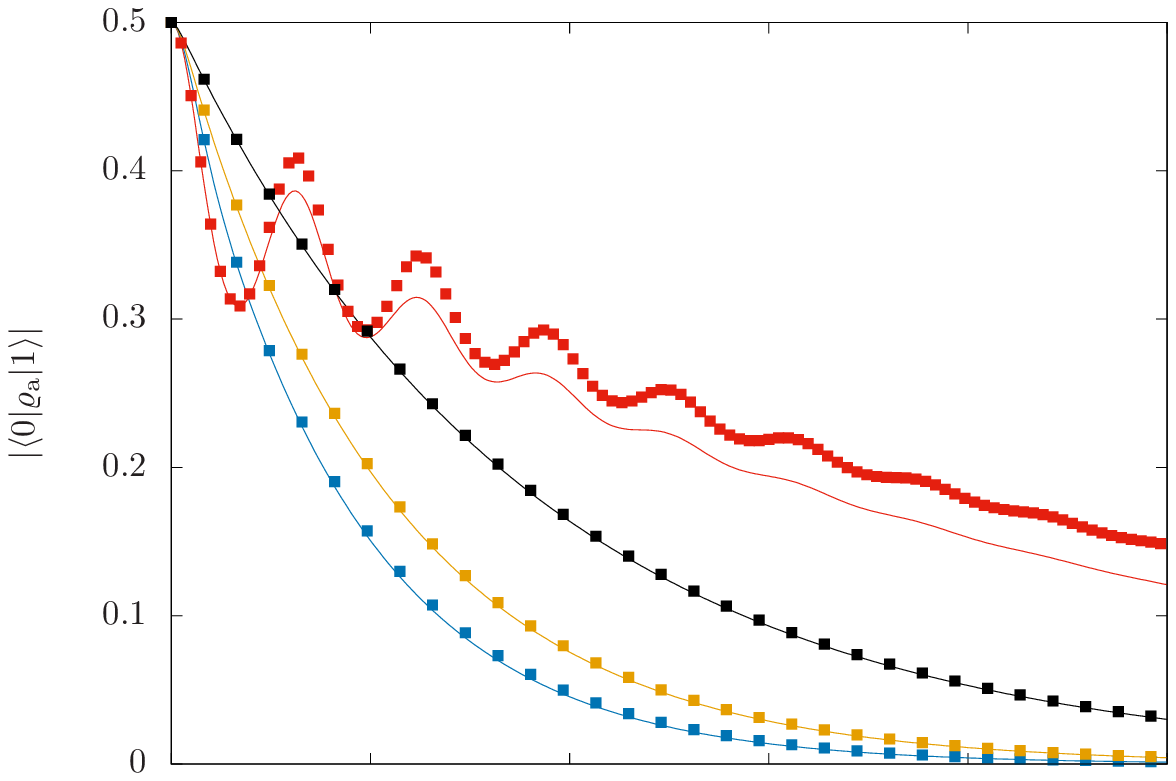}

\vspace{-5mm}

\includegraphics[width=0.5\textwidth]{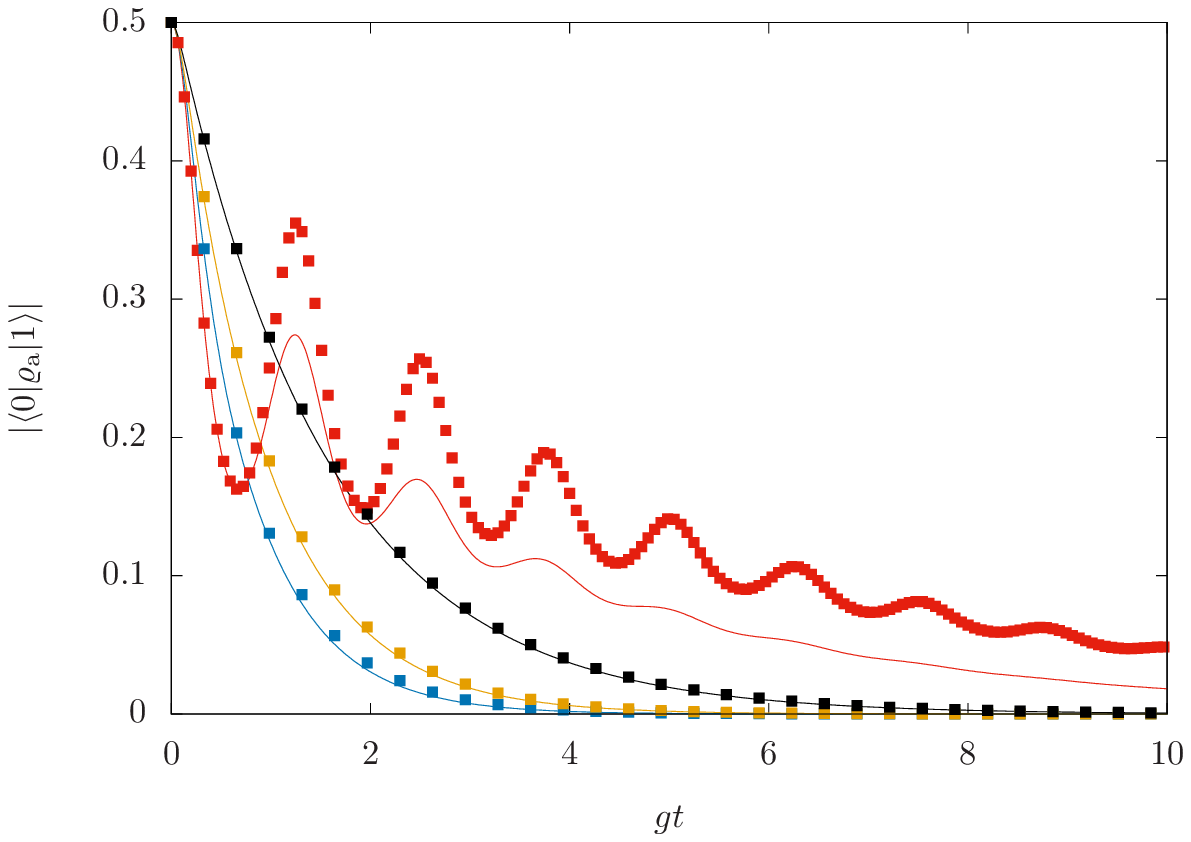}
\caption{(Color online) The coherence for an eigenstate of $\sigma_x$ as 
initial state for the dephasing coupling between qubit and near environment. 
Solid lines show results for the QO dissipation model, square points for 
the DH model. The color coding identifies the different dissipation rates: 
$\kappa/g = 1.0$ (red), $10.0$ (blue), $20$ (yellow), and $40$ (black). For the 
DH model, all rates have been reduced by one half. 
Upper panel: Temperature $\bar n = 1.0$, dephasing coupling $g=0.1$. Middle 
panel: same temperature but larger coupling, $g=0.2$. Lower panel: larger
coupling $g=0.2$ and larger temperature $\bar n = 3.0$. }
\label{f:EJ:Dephasing}\end{figure}

In Fig.~\ref{f:EJ:Dephasing}, we show our results for the QO and the DH
dissipation model. Results for the CL model have been analyzed also, but they
are not shown as they agree very well with the QO model (in the regime 
$\kappa \le 1$, where the CL model is applicable).

As explained above, the level spacing in the intermediate system determined by
$w_\rme$ is now important, providing an additional energy scale. If the 
stabilization effect would work similarly as in the random matrix cases, we 
would expect to find the effect for small values of $g$ and $\kappa/g \gg 1$. 
This is clearly not the case, as can be seen in the upper panel, where 
$g= 0.1$. Up to $\kappa/g > 10$ no stabilization effect can be observed, and
only between $\kappa/g = 20$ (yellow lines) and $\kappa/g = 40$ (black line)
some stabilization effect may seem to set in. 

What is even more surprising, for the larger values of $g$ (\textit{e.g.}
$g=0.2$ as in the middle panel), the stabilization effect seem to set in even
earlier. There we find the turnover between $\kappa/g = 10$ and 
$\kappa/g = 20$. This is confirmed in the lower panel, where the value for 
$g$ is the same, but the temperature was increased to $\bar n = 3.0$. In this
case, the turnover is again observed between $\kappa/g = 10$ and 
$\kappa/g = 20$. Though note that the overall decoherence rate is increased 
due to the higher temperature.

\section{\label{C} Conclusions}

We introduced the depolarizing heat bath (DH) as a minimal dissipation model, 
which fulfills all the basic requirements of a heat bath. We use this model to 
study the relaxation of a central system, coupled via an intermediate system 
(the near environment) to that heat bath. At strong dissipation, \textit{i.e.} 
if the dissipation rate is larger than the coupling between central system and
near environment, the DH model seems to become universal in the following
sense:

Consider a bipartite system, consisting of a central system and a near 
environment, coupled to an outer heat bath. Assume that its relaxation process 
is described by some quantum Markovian master equation, with dissipation terms 
which involve the near environment degree(s) of freedom, only. Assume further, 
that we study the relaxation of the central system, when the initial state is
a product state with the near environment in thermal equilibrium with the heat 
bath. Then we conjecture that in the strong coupling case, the dynamics of the 
central system can be described equally well with the help of the DH 
dissipation model.

Previously, this has been shown for bipartite and tripartite random matrix 
models~\cite{MGS15,GMS16RTSA} in the limit of infinite temperature. 
Here, we give numerical evidence for a bipartite model consisting of a 
two-level system and a harmonic oscillator mode, with Jaynes-Cummings or 
dephasing coupling, and using two different fundamental dissipation models 
(the quantum optical and the Caldeira-Leggett model) to describe the coupling 
between the oscillator mode and a heat bath.

A bipartite quantum system appears naturally, in the pseudo-mode
theory developped in~\cite{Garra97}, and generalizations to the case of 
multiple modes~\cite{Dal01,Rod12}, which lead to Markovian master equations on
the expense of enlarging that part of the system which is described by 
Hamiltonian dynamics. We may thus divide the system into two parts: the 
``central system'' which contains those degrees of freedom were are interested
in, and the remaining (added) degrees of freedom, we call the ``intermediate 
system''. We found numerical evidence that such dissipative bi-partite 
systems, may be described by a universal dissipation model (the depolarizing
heat bath), when the coupling to the external heat bath (dissipation) is
sufficiently strong. 

In that limit the coupling to the heat bath may be considered as an ideal
thermal contact and thus, a setup as described here, may be useful for the 
study of quantum thermal machines, as it may provide an alternative method for
defining or determining thermodynamic quantities such as heat and work.

\begin{acknowledgments}
We are greateful for very fruitful discussions with A. Eisfeld, P.~C. L\' opez,
J.~M. Torres and T.~H. Seligman. ABR and TG acknowledge the hospitality of the 
Max-Planck-Institut for the Physics of complex systems, as well as the Centro 
Internacional de Ciencias, where many of these discussions took place.
\end{acknowledgments}

\appendix

\section{\label{aT} Thermal equilibrium states of the harmonic oscillator}

As far as the temperature in the diffusion term is concerned,
note that for the harmonic oscillator state $w_T$ in thermal equilibrium, it 
holds
\begin{align}
w_T &= \sum_{n=0}^\infty \varrho_{nn}\; |n\ra\la n| \; , \qquad
	\varrho_{nn} = \frac{\rme^{-bn}}{Z}\; , \quad\text{where}\notag\\
 b &= \frac{\hbar w_\rme}{k_{\rm B} T} 
\quad\text{and}\quad
Z = \sum_{n=0}^\infty \rme^{-bn} = \frac{1}{1- \rme^{-b}} \; .
\end{align}
Therefore,
\begin{align}
\bar n &= \frac{1}{Z}\sum_{n=0}^\infty n\; \rme^{-bn} 
   = \frac{-\partial_b\, Z}{Z} = -\partial_b\; \ln(Z)
   = \partial_b\; \ln(1 - \rme^{-b})\notag\\
 &= \frac{\rme^{-b}}{1- \rme^{-b}} 
  = \frac{\rme^{-b/2}}{\rme^{b/2} - \rme^{-b/2}} 
  = \frac{{\rm coth}(b/2) -1}{2}\notag\\
&\Rightarrow \quad {\rm coth}\Big (\frac{\hbar w}{2k_B T}\Big ) 
  = 2\bar n + 1\; .
\label{aT:HOtempbarn}\end{align}

\section{\label{aC} Caldeira-Leggett model in original variables} 

The original CL-model~\cite{CalLeg83,BrePet02}, has the following master 
equation (we use the notation of Ref.~\cite{PLG17})
\[ \rmi\hbar\, \partial_\tau\varrho = \hbar w_\rme\, [a^\dagger a , \varrho] 
  + \gamma\; [\hat X,\{ \hat P, \varrho\}] 
  - 2\rmi\gamma\, \frac{m_\rme k_{\rm B} T}{\hbar}\, 
      [\hat X , [\hat X, \varrho]] \; , \]
where $\tau$ denotes the original time in physical units. Replacing the 
physical position and momentum operators by adimensional ones, \textit{i.e.}
$\hat X= \sqrt{\hbar/(m_\rme w_\rme)}\, \hat x$ and 
$\hat P= \sqrt{\hbar\, m_\rme w_\rme}\, \hat p$ we find
\[ \rmi\, \partial_\tau\varrho = w_\rme\, [a^\dagger a , \varrho] 
  + \gamma\; [\hat x,\{ \hat p, \varrho\}] 
  - \rmi\gamma\, \frac{2k_{\rm B} T}{\hbar w_\rme}\, 
      [\hat x , [\hat x, \varrho]] \; . \]
In order to deal gracefully with the low temperature regime, we 
follow the original work~\cite{CalLeg83}, and consider the ratio 
$2k_{\rm B} T/(\hbar w_\rme)$ as the high temperature limit of 
\[ {\rm coth}\Big ( \frac{\hbar w}{2k_{\rm B} T}\Big ) = 2\bar n + 1\; , \]
where $\bar n$ is the average number of excited modes of the harmonic 
oscillator at thermal equilibrium. Finally, we switch to the adimensional time 
$t= w_\rme\tau$ and thus obtain:
\begin{align}
\rmi\partial_t\, \varrho &= w_\rme\, [\hat a^\dagger\hat a, \varrho]
   + \gamma\, [\hat x,\{ \hat p,\varrho\}] \notag\\
&\qquad\qquad\qquad -\rmi\, \gamma\, (2\bar n +1) \; 
   [\hat x, [\hat x,\varrho]] \; ,
\end{align}
where $\gamma$ is the classical damping rate, i.e. the system loses energy 
with that rate.  Comparing to the master equation with the general form, 
Eq.~(\ref{M:MaEc}) in Sec.~\ref{M}, to find
\begin{align}
\mathcal{D}_{\rm CL}[\varrho] = 2\rmi\, [\hat x, \{ \hat p,\varrho\} ]
   +2\, (2\bar n +1)\, [\hat x, [\hat x,\varrho]] \; .
\label{aC:CalLegDiss}\end{align}

\bibliographystyle{unsrt}
\IfFileExists{/home/alexul/Bib/JabRef-Deli.bib}
   {\bibliography{/home/alexul/Bib/JabRef}}
   {\bibliography{/home/gorin/Documentos/Bib/JabRef-Deli}}

\end{document}